\documentclass[aps,prapplied,twocolumn,superscriptaddress]{revtex4-2}
\usepackage{graphicx}
\usepackage{gensymb}

\begin{document}

\title{Wide-aperture layered-sheet Faraday isolator}

\author{R.~ Kononchuk}
\affiliation{Department of Physics and Astronomy, University of Texas at San Antonio, San Antonio, TX 78249, USA}

\author{C.~Pfeiffer}
\affiliation{Air Force Research Laboratory, Sensors Directorate, Wright-Patterson AFB, OH 45433, USA}

\author{I.~Anisimov}
\affiliation{Air Force Research Laboratory, Sensors Directorate, Wright-Patterson AFB, OH 45433, USA}

\author{N.I.~Limberopoulos}
\affiliation{Air Force Research Laboratory, Sensors Directorate, Wright-Patterson AFB, OH 45433, USA}

\author{I.~Vitebskiy}
\affiliation{Air Force Research Laboratory, Sensors Directorate, Wright-Patterson AFB, OH 45433, USA}

\author{A. A.~Chabanov}
\affiliation{Department of Physics and Astronomy, University of Texas at San Antonio, San Antonio, TX 78249, USA}

\date{\today}

\begin{abstract}
We introduce the first conceptual design of a free-space thin-sheet isolator with unlimited aperture and the possibility of a broadband omnidirectional rejection of the backward propagating light. The proposed design involves a multilayered resonant cavity incorporating subwavelength magnetic layers, dichroic nanolayers, and an optional metallic nanolayer. The cavity resonance enhances the Faraday rotation produced by the subwavelength magnetic layers, while providing nearly total absorption of the backward-propagating light by the dichroic nanolayers. The latter is a necessary and the most challenging condition for a thin-sheet isolator with unlimited aperture to function. The (optional) metallic nanolayer provides rejection of the obliquely incident light, which otherwise would be partially transmitted in either direction. Our numerical simulations and quasi-optical measurements at millimeter-wave frequencies illustrate how the key elements of the layered-sheet isolator work. Our approach can be scaled down to long- and mid-infrared wavelengths.

\end{abstract}

\maketitle

\section{Introduction}

Optical isolators are nonreciprocal devices transmitting light in only one direction \cite{optics_book}. They are indispensable for eliminating adverse effects of back reflection in optical systems. A typical free-space isolator consists of a 45-degree Faraday rotator placed between two polarizers making a 45-degree angle with each other, as schematically shown in Fig.~1. A properly polarized input light propagating in the forward direction, passes through the front polarizer P1, the Faraday rotator 45FR, and then the back polarizer P2, as shown in Fig.~1a. By contrast, the backward-propagating light, after passing the back polarizer P2 and the Faraday rotator 45FR, will be blocked by the front polarizer P1, as shown in Fig.~1b. This simple isolation scheme essentially requires that the backward-propagating light is either totally absorbed by the polarizer P1 or diverted away from the system. Indeed, if the backward-propagating light in Fig.~1b were even partially reflected from the polarizer P1 towards the Faraday rotator, it would bounce back and force between the two polarizers, each time passing through the Faraday rotator and acquiring an additional 45-degree rotation. After one round trip, the light would acquire an additional 90-degree polarization rotation and thus would not be blocked by the polarizer P1, resulting in isolation failure. This presents a major challenge in realizing a thin-sheet isolator, because all existing sheet polarizers (such as absorptive film polarizers, metal nanoparticles-embedded glass plates, wire grids, etc.) are at least partially reflective. In the existing free-space isolators, the problem of polarizer reflectivity is dealt with by either using beam-splitting polarizers or by tilting the sheet polarizers at an angle to the light beam \cite{Wylde2009,Hunter2007}. Neither of these approaches, however, is applicable to a thin-sheet isolator. Indeed, if the free-space isolator is a thin multilayer (as is the case here), one can only tilt the entire multilayered structure, not just the polarizer layers. Tilting the entire multilayer, however, would not prevent the backward-propagating light from bouncing between the sheet polarizers, causing the isolation failure. It thus appears that after entering the sheet isolator, the backward-propagating light must be completely absorbed by the isolator. This strict requirement is unique to the case of a sheet isolator with unlimited aperture, and its fulfillment is the main focus of our study.
\begin{figure}
\includegraphics[width=8.6cm]{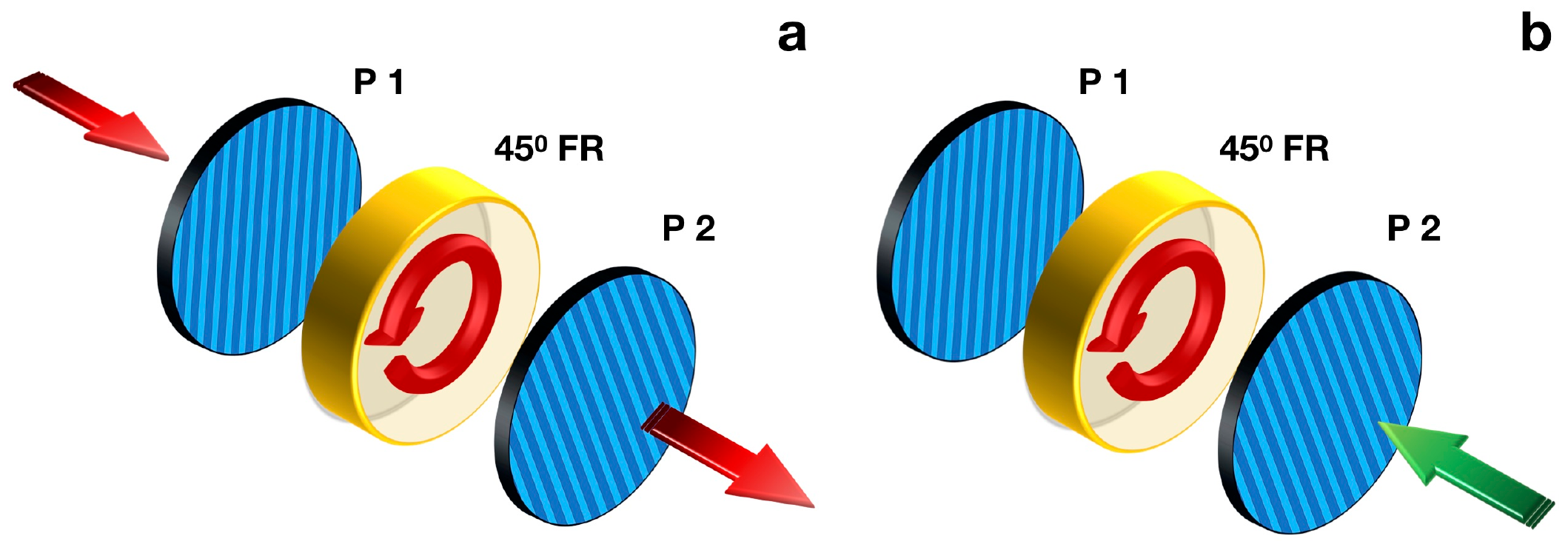}
\caption{ Schematics and operating principle  of the optical isolator composed of a 45-degree Faraday rotator (45FR) positioned between two polarizers (P1 and P2) making a 45-degree angle with each other. (a) forward-propagating wave (red arrow) passes through; (b) backward-propagating wave (green arrow) is blocked by the isolator.}
\label{figure1}
\end{figure}

In the next two Sections, we describe the design and experimental realization of layered-sheet (LS) polarizers providing nearly perfect resonant absorption of the light with unwanted linear polarization, while not affecting the light with the desired linear polarization. According to Fig.~1, a pair of such LS polarizers in combination with a thin-sheet Faraday rotator will act as a sheet isolator with unlimited aperture. Away from the cavity resonance, the proposed resonant LS polarizers, as well as the entire sheet isolator, are highly reflective regardless of the input light polarization. The latter implies that away from the cavity resonance, the isolator reflects both the forward- and the backward-propagating input light.

In Section 2 we present two qualitatively different designs of resonant LS polarizers compatible with the sheet isolator. The first design is symmetric -- it provides the total absorption of the unwanted polarization regardless of the direction of incidence. The second (preferable) design is highly asymmetric -- it provides the total absorption of the unwanted polarization for one direction of incidence, while providing a nearly perfect reflection of the unwanted polarization for the opposite direction of incidence. In the case of sheet isolator, the absorptive sides of the polarizers must face the Faraday rotator, while the opposite (outer) sides of the polarizers can be reflective. Our designs involve high-Q resonant cavity with embedded dichroic nanolayers (metal-grid polarizers). Taken out of the cavity, the dichroic nanolayers are mostly reflective for the unwanted light polarization and, therefore, could not be used as components of the sheet isolator. Both, symmetric and asymmetric designs are based on the same high-Q resonant cavity; the only difference is in the number and the location of the dichroic nanolayers inside the cavity. We have also fabricated and successfully tested such polarizers at millimeter-wave (MMW) frequencies. The results of the quasi-optical MMW measurements are described in Section 3.

The above solution for the sheet isolator with unlimited aperture involves at least two resonant cavities (one for each of the two polarizers in Fig.~1) with perfectly matched resonant frequencies. In Section 4, we put forward a qualitatively different design of the LS isolator. The idea is to incorporate the magnetic and dichroic layers in a single resonant cavity, to avoid any impedance mismatch between the three constitutive components of the isolator in Fig.~1. The role of the high-Q dielectric cavity in this case is two-fold. Firstly, it provides a nearly total resonant absorption of the backward-propagating light by the dichroic nanolayers, while not affecting the forward-propagating light. Secondly, the cavity enhances the nonreciprocal circular birefringence associated with the subwavelength magnetic layers, thereby drastically reducing the total thickness of the multilayered structure. Furthermore, a metallic nanolayer can be added to multilayered cavity to achieve omnidirectional rejection of the light incident on the back surface of the sheet isolator, following the scheme of \cite{Kyle2013,Kyle2014}. The location of metallic nanolayer should coincide with an asymmetric nodal plane of the resonant field distribution. According to \cite{Kyle2014}, at normal propagation, such a metallic nanolayer is ``invisible’’ for the resonant mode inside the cavity. However, at oblique incidence, the nodal plane of the electric field distribution shifts away from the metallic nanolayer and the entire multilayered structure becomes highly reflective \cite{Kyle2014}. In this paper, though, we focus on the case of normal incidence.

The numerical simulations of the scattering parameters for the LS polarizers and the integrated LS isolator, as well as the quasi-optical proof-of-concept measurements, have been carried out in the W-band (75-110 GHz). The presented approach, though, is highly scalable and can be replicated for any frequency range from microwave to mid-infrared. 

\section{Layered-sheet polarizers}

The design of the LS polarizers involves dichroic nanolayers incorporated into a resonant multilayer dielectric cavity. In our numerical simulations using 4$\times$4 Transfer-Matrix formalism, the dichroic nanolayers have a highly anisotropic sheet resistance: ${\cal R}_x =\infty$ and ${\cal R}_y = 2.54$ $\Omega$/sq. At MMW frequencies, the dichroic nanolayers fully transmit the $x$-polarization (hereinafter, the TM polarization) component of the incident wave in the $z$-direction while mostly reflecting the $y$-polarization (hereinafter, the TE polarization) component. The corresponding transmittance and reflectance values at a frequency of 95 GHz are $T_x = 1$, $R_x = 0$ and $T_y = 1.79\times 10^{-4}$, $R_y = 0.973$. Thus, the high value of $R_y$ would not allow the use of the stand-alone dichroic nanolayers as polarizers in the isolator in Fig.~1.

The multilayer cavity is shown in Fig.~2a, and it is designed to operate at 95 GHz. It consists of quarter-wave layers of sapphire ($H$, dark blue) and quartz ($L$, light blue), with the corresponding refractive indices $n_H=9.5$ and $n_L=3.8$ and negligible absorption in the W-band. The transmittance and reflectance spectra of the multilayer cavity are shown in Fig. 2b. The half-wave defect layer in the middle of the periodic layered structure provides a quasi-localized mode associated with the resonant transmittance at the mid-gap frequency  $f_0=95$ GHz. The spatial intensity distribution of the electric field component within the multilayer at the resonance frequency $f_0$ is shown in Fig. 2c. Since the half-wave defect layer has the lower refractive index, the resonant field distribution has an antinodal plane in the middle of the defect layer, where the amplitude of the quasi-localized resonant mode reaches its maximum value. In the absence of the dichroic layers, the results presented in Figs.~2b and 2c are polarization independent. These results will still apply to the TM polarization, after the dichroic layers have been incorporated into the multilayer. The inclusion of the dichroic nanolayers in the multilayer cavity provides a broadband, omnidirectional and strongly enhanced transmittance suppression for the TE polarization. What happens to the TE polarization strongly depends on the number and the location of the dichroic nanolayers in the cavity.

\begin{figure}
\includegraphics[width=8.6cm]{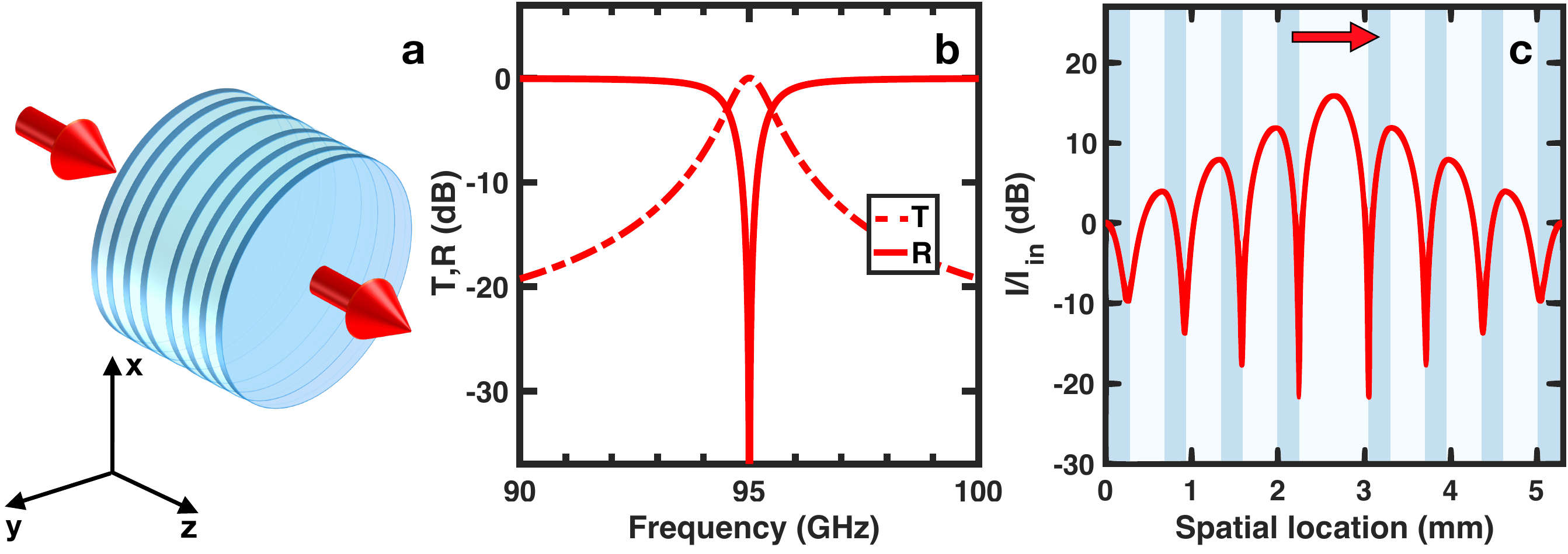}
\caption{(a) Schematics of the resonant multilayer cavity of quarter-wave sapphire (deep blue) and quartz (light blue) layers, utilized as a host structure for the LS polarizers in Figs.~3--5; (b) simulated  spectral transmittance (dashed line) and reflectance (solid line), and (c) spatial intensity distribution, $|E(z)|^2$, at the cavity resonance frequency 95 GHz. The plots (b) and (c) are polarization independent. In the case of LS polarizers, they still apply to the TM polarization.}
\label{figure2}
\end{figure}

We start with the symmetric LS configuration in Fig.~3a, where the only dichroic nanolayer coincides with the symmetry plane of the cavity. Our simulations show that the resonant transmittance of the TE polarization is $\sim 10^{-7}$ (as seen in Fig.~3c), which is three orders of magnitude lower than for the stand-alone dichroic layer. The reflectance approaches unity (as seen in Fig.~3b), while the absorptance, $A=1-T-R$, drops an order of magnitude lower than that of the stand-alone dichroic layer. The physical reason for the reduced absorption is that the TE polarization is mostly reflected from the Bragg-reflector part of the LS polarizer and that the electric field intensity at the dichroic layer is only $\sim 10^{-3}$ of the incident wave intensity, as seen in Fig.~3c. The decreased absorption can drastically reduce the absorption-related heating and, thereby, significantly increase the power-handling capability of the LS polarizer. However, its high reflectance for the TE polarization makes it impossible to use it as a component of the LS isolator.

\begin{figure}
\includegraphics[width=8.6cm]{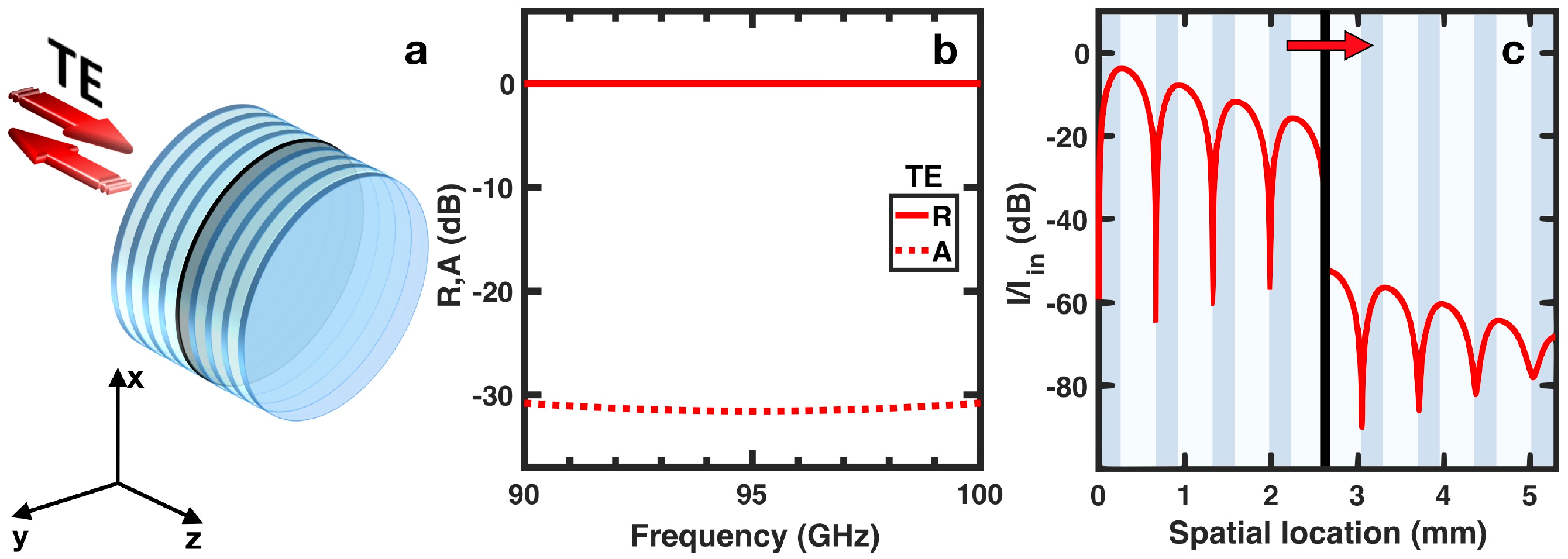}
\caption{(a) Schematics of the reflective LS polarizer involving a dichroic nanolayer (black) incorporated in the host multilayer structure of Fig.~2a; (b) simulated spectral reflectance (solid line) and absorptance (dotted line), and (c) spatial intensity distribution, $|E(z)|^2$, at the cavity resonance frequency 95 GHz, for the TE polarization in the forward propagation direction.}
\label{figure3}
\end{figure}

If, on the other hand, the dichroic layer is positioned at the nodal plane of the resonant electric field distribution, resonant absorption of the TE polarization can be significantly enhanced. Although the absorptance of an ultrathin homogeneous layer is known not to exceed 0.5 (see, for example, \cite{Kasemo2010}), this absorptance limit can be overcome in a thin film with an intrinsic resonance \cite{Zheludev2012}, such as a plasmonic resonance, in a metasurface \cite{Huang2012}, or by using the Salisbury screen \cite{Salisbury1952,Fante1988}, which is a highly reflective surface positioned a quarter-wave distance behind an absorbing layer. Perfect absorption can also be achieved by critical coupling of a lossy thin layer to a resonator \cite{Tischler2006,Piper2014,Fan2014} or ideal impedance matching between a lossy slab and the free space \cite{Nefedov2013,Baranov2015}.

In the LS configuration in Fig.~4a, we utilize, as a reflecting screen, the second dichroic nanolayer positioned a quarter-wave distance from the first one, resulting in nearly perfect absorption of the TE polarization at the resonance frequency. The third dichroic nanolayer is added for symmetric operation, so that the LS polarizer equally absorbs the TE-polarization component of the incident wave in either direction. The resonant transmittance for the TE polarization is $\sim 10^{-8}$ (as seen in Fig.~4c), implying a huge enhancement of the polarization extinction ratio. The reflectance and absorptance spectra are shown in Fig.~4b. The nearly perfect absorptance is narrowband, due to its resonant nature. Away from the resonance frequency, the LS polarizer is highly reflective, regardless of the incident polarization. Thus in the vicinity of the cavity resonance, the LS polarizer of Fig.~4a can be used as a component of the LS isolator. 

\begin{figure}
\includegraphics[width=8.6cm]{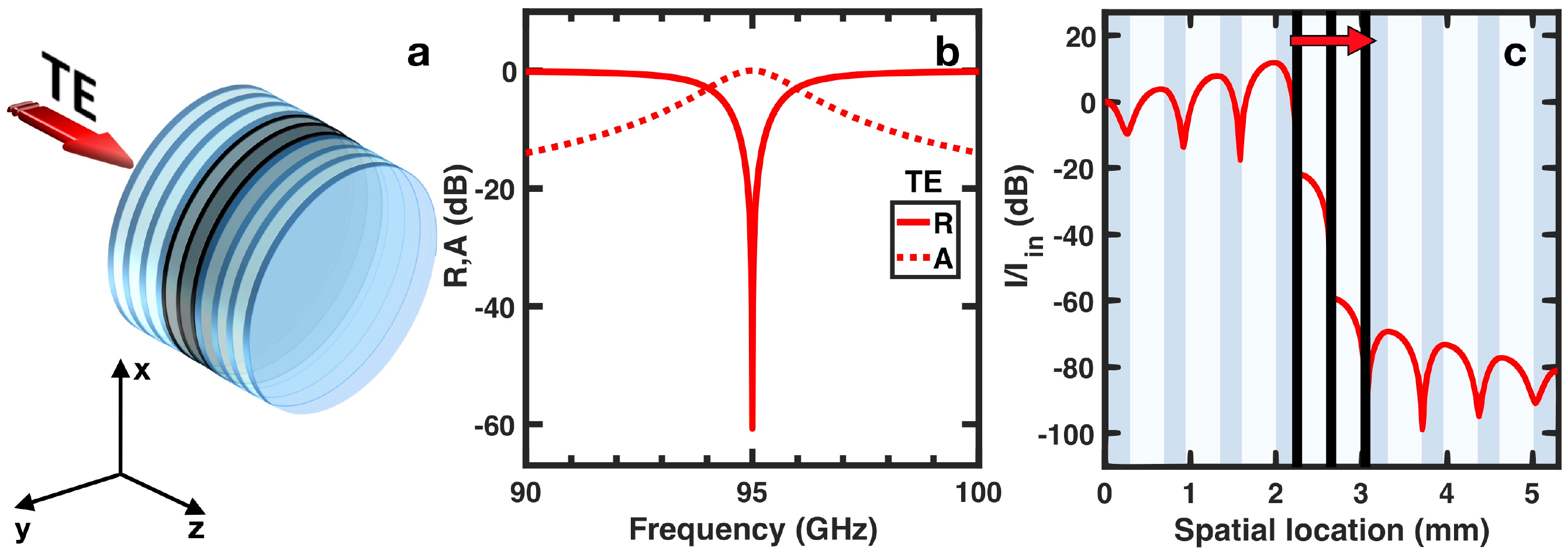}
\caption{(a) Schematics of the absorptive LS polarizer involving three dichroic nanolayers (black) incorporated in the host multilayer structure of Fig.~2a; (b) simulated spectral reflectance (solid line) and absorptance (dotted line), and (c) spatial intensity distribution, $|E(z)|^2$, at the cavity resonance frequency 95 GHz, for the TE polarization in the forward propagation direction.}
\label{figure4}
\end{figure}

Furthermore, if the left dichroic nanolayer of the LS polarizer in Fig.~4a is removed, the layered structure becomes asymmetric, as shown in Fig.~5a.  Despite the structural asymmetry, the transmittance of the LS polarizer remains identical for the opposite incident directions, due to the reciprocity principle. In particular, the resonant transmittance for the TE polarization is less than $10^{-7}$ (as seen in Fig.~5c), still implying an enormous enhancement of the polarization extinction ratio. By contrast, the reflectance and absorptance are extremely asymmetric about the direction of incidence, as seen in Fig.~5b. At the frequency of the cavity resonance, the wave with the TE polarization traveling from left to right is almost totally reflected by the asymmetric LS polarizer, while the wave with the TE polarization traveling in the opposite direction is almost totally absorbed, primarily by the right dichroic nanolayer. Note that the Salisbury screen, which is inherently asymmetric, can also produce, at the resonance frequency, strongly asymmetric absorption/reflection of the incident wave \cite{Salisbury1952}. It will, however, have the polarization extinction ratio and power limitation of a common dichroic-sheet polarizer. By comparison, in the case of the asymmetric LS polarizer of Fig.~5a, the polarization extinction ratio and power-handling capability are greatly enhanced.

\begin{figure}
\includegraphics[width=8.6cm]{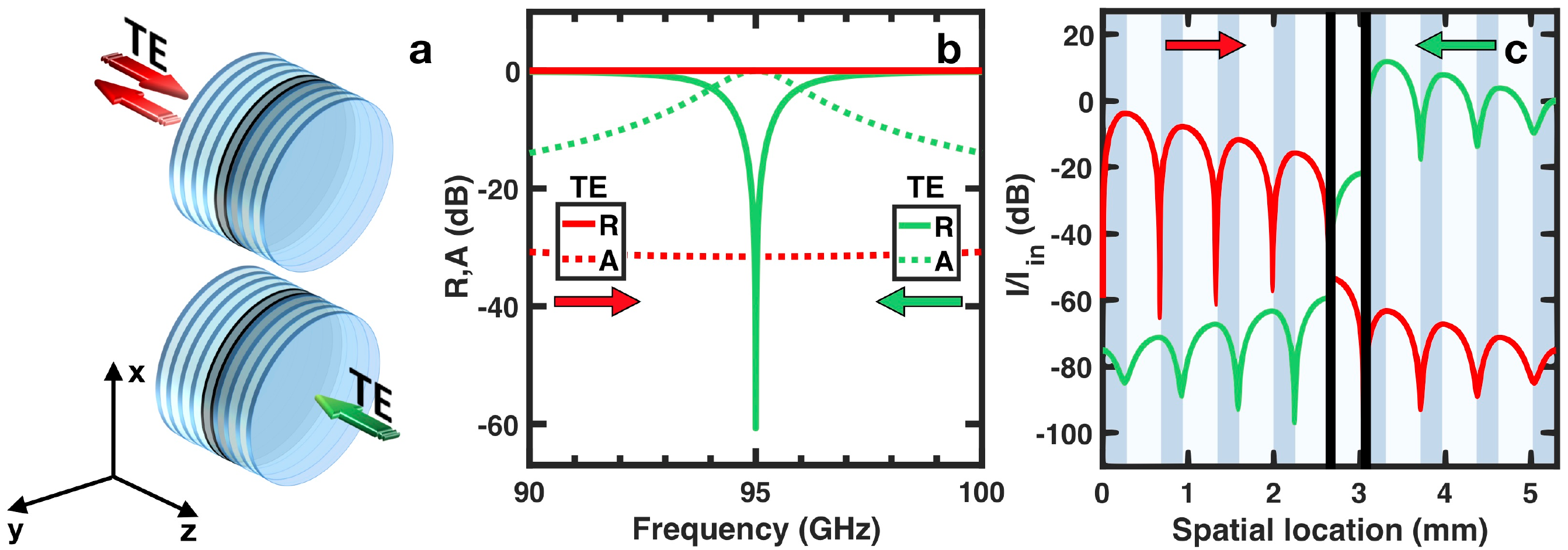}
\caption{(a) Schematics of the asymmetric absorptive/reflective LS polarizer involving two dichroic nanolayers (black) incorporated in the host multilayer structure of Fig.~2a; (b) simulated spectral reflectance (solid line) and absorptance (dotted line), and (c) spatial intensity distribution, $|E(z)|^2$, at the cavity resonance frequency 95 GHz, for the TE polarization in the forward (red) and backward (green) propagation directions.}
\label{figure5}
\end{figure}

\section{Quasi-optical measurements of the layered-sheet polarizer}

The proof-of-concept measurements of the LS polarizers introduced in Section 2 have been carried out in the W-band. The asymmetric LS polarizer was fabricated using 5-cm-diameter C-cut sapphire and quartz wafers and aluminum (Al) wire grids to produce the layer stack ($H_1L_1$)$^2$($H_1L_2$)($GH_2$)$^2$($L_2H_1$)($L_1H_1$)$^2$, where $L_1$ and $L_2$ are 404-$\mu$m and 442-$\mu$m thick quartz, respectively, $H_1$ is a 256-$\mu$m thick sapphire, and $(GH_2)$ is a 40-nm thick Al wire grid with a duty cycle of 0.3 and period of 20 $\mu$m, deposited on a 246-$\mu$m thick sapphire wafer. The layer structure was mainly determined by the wafers on hand, and the wire grid was designed to fit to the fabricated multilayer.

First, quasi-optical  measurements of the wire grid on a double-thick sapphire $(GH_2^2)$ were carried out with the use of Gaussian beams (beam waist diameter of $\sim 0.5$ cm) normally incident to the grid. The S-parameters of $(GH_2^2)$ were measured as a function of frequency with the use of a vector network analyzer. No significant difference was noticed, within experimental error, between the measurements for the two (opposite) incident directions. The transmittance and reflectance spectra of $(GH_2^2)$ for the TM and TE polarizations are shown in Figs.~6a and 6b, respectively. The sharp dip in the reflectance for the TM polarization at 94.1 GHz indicates the position of the Fabry-P\'{e}rot resonance of the substrate. At the resonance frequency, the polarization extinction ratio of $(GH_2^2)$ is 33 dB. For the TE polarization, the reflectance is 0.86, implying that $(GH_2^2)$ has a significant absorptance of 0.14.

\begin{figure}
\includegraphics[width=8.6cm]{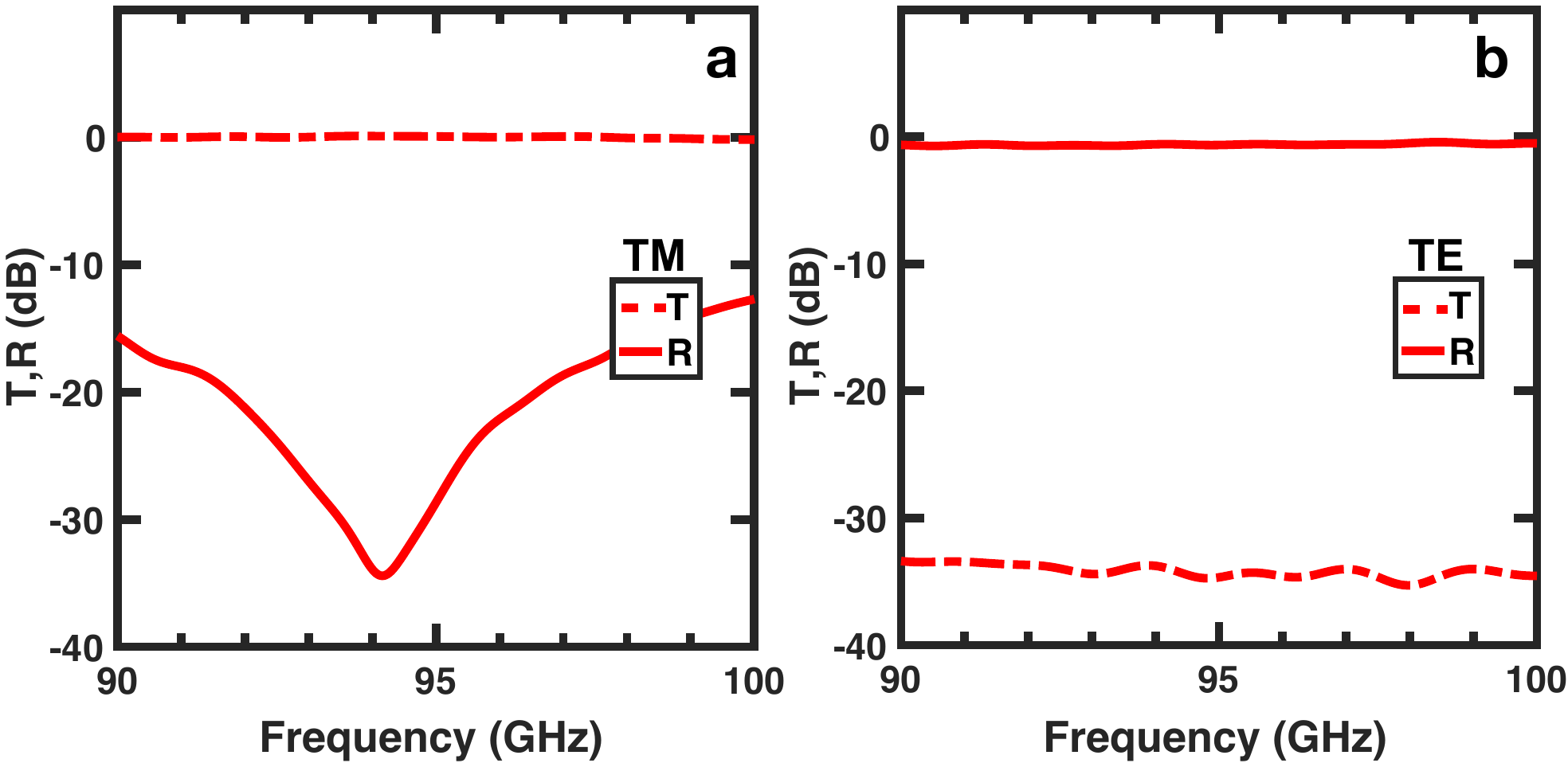}
\caption{Quasi-optical spectral measurements of the aluminum wire grid on sapphire: transmittance (dashed line) and reflectance (solid line) of the (a) TM- and (b) TE-polarized Gaussian beam.}
\label{figure6}
\end{figure}

The S-parameter measurements were then performed on the layer stack. For the TM polarization, the difference between the measurements for the two incident directions was negligible, similarly to the case of $(GH_2^2)$. The transmittance and reflectance spectra for the TM polarization are shown in Fig.~7a. The layer stack has a nearly perfect resonant transmittance (with a resonance Q-factor of 92) at 94.4 GHz and high reflectance away from the cavity resonance, indicating that the TM polarization does not interact with the wire grids. For the TE polarization, the transmittance is suppressed down to -50 dB within the entire band gap, being identical for the two incident directions (Figs.~7b and 7c). By comparison, the absorptance and reflectance are highly asymmetric about the direction of incidence. For the Gaussian beam incident from the left (Fig.~7b), the reflectance is 0.99 in the vicinity of the resonance frequency, implying that the absorptance is less than 0.01. By contrast, for the Gaussian beam incident from the right (Fig.~7c), the reflectance has a sharp dip of -38 dB at 94.03 GHz, indicating that the TE polarization is almost totally absorbed at the cavity resonance. The red-shift of the resonance frequency for the TE polarization is attributed to parasitic reactances of the wire grids, which effectively increase the cavity thickness. According to our simulations, this effect can be reduced by decreasing the period of the wire grid. Notwithstanding the different frequencies of the cavity resonance for the TM and TE polarizations (in Figs.~7a and 7c), our experimental results check well with those presented in Fig.~5b. The polarization extinction ratio as high as 50 dB and strong asymmetry of the absorptance/reflectance about the direction of incidence, in the vicinity of the resonance frequency, have both been experimentally demonstrated. In addition, the extinction ratio of the LS polarizer was not adversely affected when it was tilted at an angle as large as 40 degrees to the incoming beam. The oblique incidence, though, resulted in a blue-shift of the resonant frequency. 

\begin{figure}
\includegraphics[width=8.6cm]{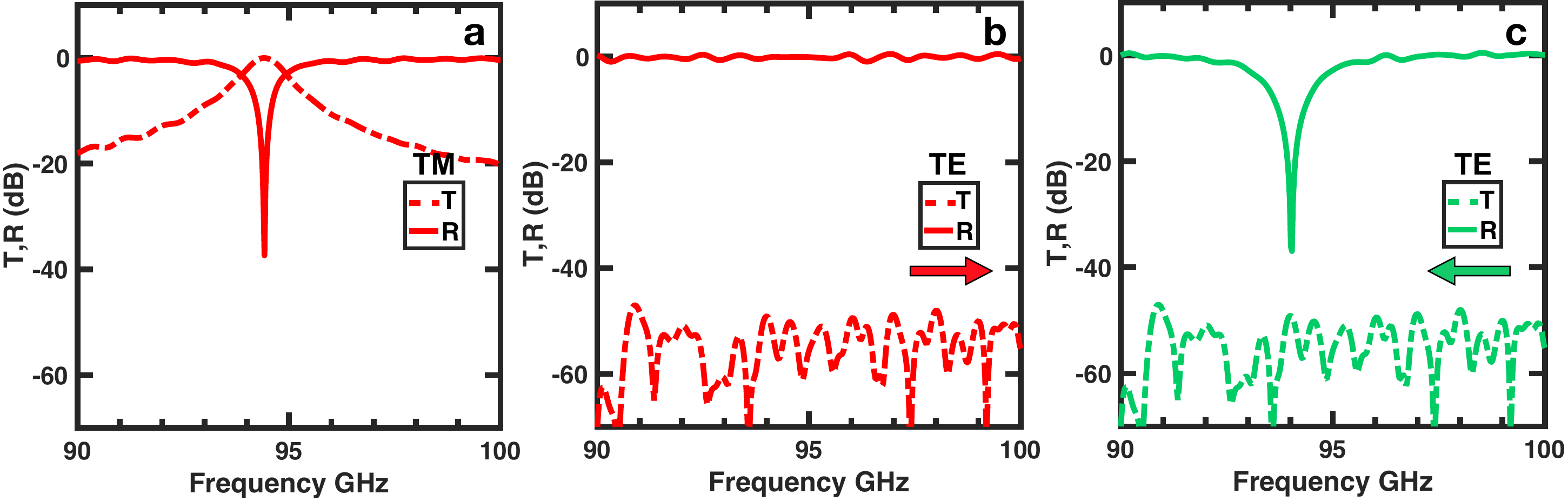}
\caption{Quasi-optical spectral measurements of the asymmetric LS polarizer: transmittance (dashed line) and reflectance (solid line) of (a) the TM-polarized Gaussian beam, (b) the TE-polarized Gaussian beam incident to the right and (c) to the left.}
\label{figure7}
\end{figure}

\section{Integrated layered-sheet isolator}

A direct way to assemble a LS isolator is by placing a thin-sheet 45-degree Faraday rotator between a pair of the absorbing LS polarizers (such as in Figs.~4a or 5a), following the scheme of Fig.~1. A thin-sheet 45-degree Faraday rotator in the form of a resonant multilayer cavity incorporating subwavelength magnetic layers was discussed for the X-band in Ref. \cite{Kyle2013}. Thus the isolator assembly can include three resonant components operating at the same resonance frequency.

Here we present a design of LS isolator based on a single multilayer cavity, which we coin `integrated LS isolator'. The integrated LS isolator, designed for the W-band, is shown in Fig.~8. It still involves magnetic and dichroic layers incorporated in a sapphire/quartz multilayer, but the polarizers cannot be identified as separate parts of the isolator. The multilayer includes two symmetric half-wave defect layers, providing a pair of resonant cavity-localized modes which enable a low-loss Faraday rotator \cite{Inoue2002}. The magnetic layers are a 11.65-$\mu$m thick strontium ferrite (SrFe$_{12}$O$_{19}$) with the permittivity $\varepsilon=36+i0.1$ \cite{Liu2009,Shalaby2013}, diagonal elements of the permeability tensor $\mu_{ii}=1$ and specific Faraday rotation $\alpha=$ 700 deg/cm \cite{Shalaby2013}, and the dichroic nanolayers are characterized by the sheet resistances ${\cal R}_x =\infty$ and ${\cal R}_y = 15.6$ $\Omega$/sq. The position of the magnetic layers is determined by the following. Note that at MMW frequencies, the nonreciprocal response is usually associated with the magnetic permeability tensor, whereas absorption is often caused by electrical conductivity of the magnetic material \cite{Gurevich_book}. This implies that the magnetic component of the electromagnetic wave is responsible for the nonreciprocal effects (such as Faraday rotation), while the electric component for the losses. Also note that at the cavity resonance, the nodal planes of the electric field component coincide with the antinodal planes of the magnetic field component, and vice versa. Therefore, when a subwavelength magnetic layer (such as strontium ferrite) is positioned at the antinodal plane of the magnetic field (also the nodal plane of the electric field), the Faraday rotation is enhanced and the losses are suppressed at the same time \cite{Kyle2013}. This approach can be utilized for any lossy magnetic material of considerable magnetic gyrotropy, provided that the gyrotropy is predominantly associated with the permeability tensor while the losses are caused by the electric conductivity. Such magnetic materials are available from microwave to long- and mid-infrared frequencies \cite{Gurevich_book,Zvezdin_book}. 

\begin{figure}
\includegraphics[width=8.6cm]{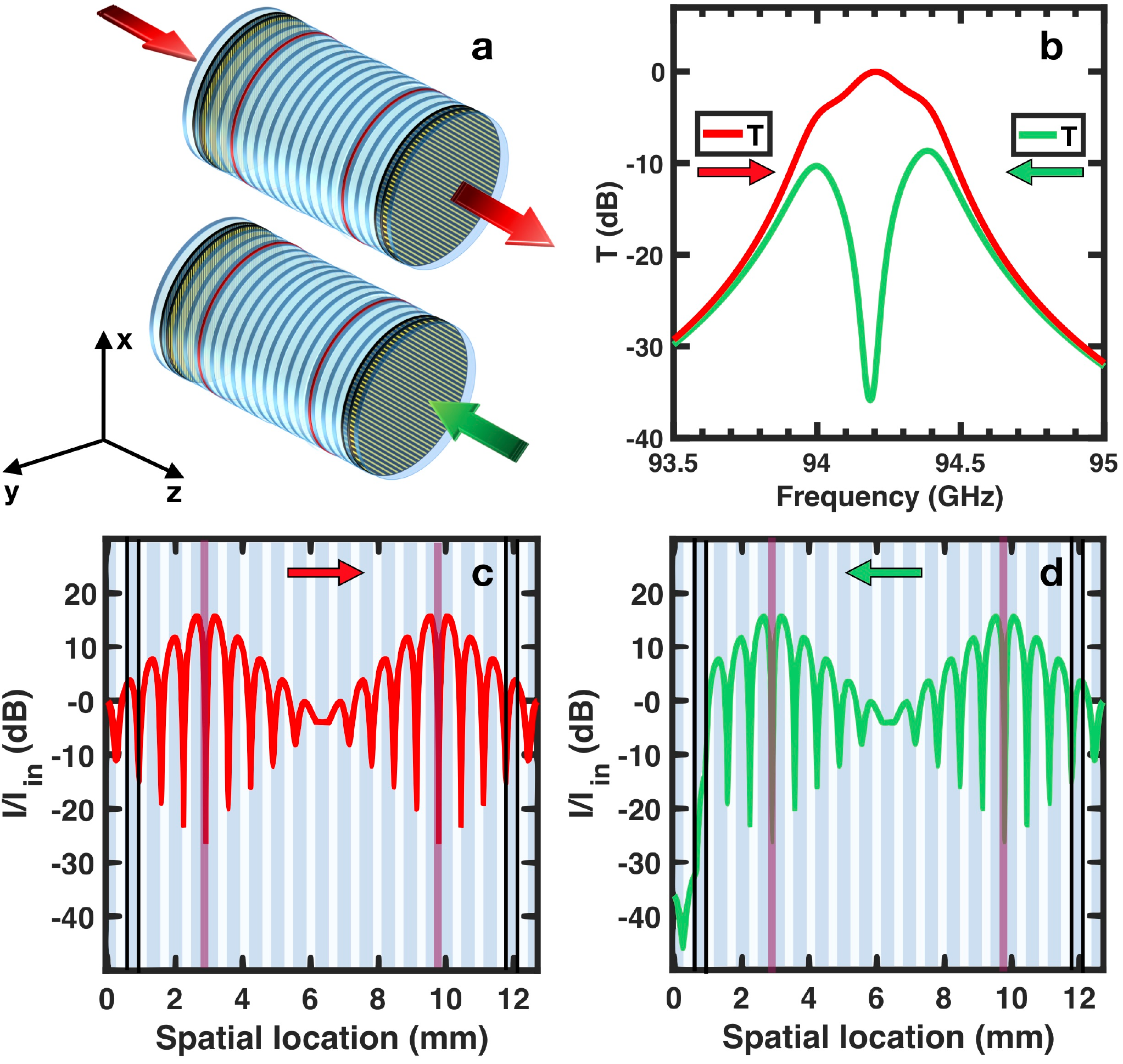}
\caption{(a) Schematics of the integrated LS isolator composed of the resonant sapphire/quartz (dark/light blue) multilayer incorporating two magnetic layers (red) and two pairs of the (aligned) dichroic nanolayers (black) making a 45-degree angle with each other; (b) simulated spectral transmittance in the forward (red) and backward (green) propagation directions in the vicinity of the cavity resonance; (c,d) the corresponding spatial intensity distributions, $|E(z)|^2$, at the resonance frequency 94.20 GHz.}
\label{figure1}
\end{figure}

The two pairs of the (aligned) dichroic nanolayers, making a 45-degree angle with each other, each act as the asymmetric polarizer. They reflect the TE-polarized component of the incoming wave and resonantly absorb the TE-polarized component of the wave passed the magnetic layers, in either direction (Figs.~8c and 8d). Note that the dichroic nanolayers can adversely affect the Faraday rotation when positioned very close to the magnetic layer. On the other hand, placing them at the edges of the multilayer degrades their polarization contrast and power-handling capability.

The integrated LS isolator, at normal incidence, achieves insertion loss of 0.04 dB and return loss of 33 dB at the resonance frequency 94.20 GHz (Fig.~8b). The return loss can be further increased by increasing the number of sapphire/quartz bi-layeres. Away from the resonance, the LS isolator is highly reflective in either direction. At oblique incidence, the degree of isolation progressively deteriorates with increasing angle of incidence, in the same way as for conventional free-space isolators. One way to address this problem is by adding a metallic nanolayer to the LS isolator \cite{Kyle2014}. The position of the metallic nanolayer should coincide with a nodal plane of the electric component of the resonant field distribution at normal incidence, to avoid a significant increase in the insertion loss \cite{Berning1957}. At oblique incidence, though, the nodal plane of the electric field distribution can shift away from the metallic nanolayer thus making the LS isolator highly reflective at any frequency. For this to happen, the nodal plane location must not coincide with the symmetry plane (if any) of the multilayered structure. This approach is discussed in greater detail in Ref. \cite{Makri2016}.

\section{Conclusion}

We have developed the wide-aperture LS isolators for use in optical and quasi-optical systems. Our design is based on a resonant multilayer dielectric cavity hosting subwavelength magnetic layers and dichroic nanolayers. The role of the cavity resonance is essential and multifold. Firstly, the resonance enhances the magnetic Faraday rotation produced by the subwavelength magnetic layers thus allowing to make the LS isolator rather thin -- its total thickness does not exceed a few wavelengths. Secondly, the resonant conditions allow the use of dichroic nanolayers by enhancing/modifying their dichroic properties. In particular, the resonant conditions provide a nearly total (resonant) absorption of the backward-propagating light by the dichroic nanolayers. The dichroic nanolayers taken out from the resonant multilayer cavity would be partially absorptive/reflective and, thus, could not be used in a sheet isolator. Finally, the resonant conditions are necessary for the formation of quasi-nodal planes in the oscillating electric field distribution. Placing a metallic nanolayer at the location of an asymmetric nodal plane results in the rejection of the obliquely incident light and rendering the LS isolator omnidirectional \cite{Kyle2014}.

\begin{acknowledgments}
This research was supported by AFOSR via grants FA9550-16-1-0058, FA8650-17-C-1026, FA9550-15-F-0001, and LRIR 18RYCOR013.
\end{acknowledgments}


\begin{thebibliography}{99}
\bibitem{optics_book}{\it Handbook of Optical Components and Engineering}, edited by K. Chang (Wiley-Interscience, New York, 2003).
\bibitem{Wylde2009}D. H. Martin and R. J. Wylde, IEEE Trans. Microw. Theory Tech. {\bf 57},  99 (2009).
\bibitem{Hunter2007}R. I. Hunter and D. A. Robertson, IEEE Trans. Microw. Theory Tech. {\bf 55},  890 (2007).
\bibitem{Kyle2013}K. Smith, T. Carroll, J. D. Bodyfelt, I. Vitebskiy, and A. A. Chabanov, J. Phys. D: Appl. Phys. {\bf 46}, 165002 (2013).
\bibitem{Kyle2014}A. A. Chabanov, K. Smith, T. Carroll, and I. Vitebskiy, in {\it 2014 8th International Congress on Advanced Electromagnetic Materials in Microwaves and Optics}  (IEEE, New York, 2014), p. 79.
\bibitem{Kasemo2010}C. H\"{a}gglund, S. P. Apell, and B. Kasemo, Nano Lett. {\bf 10}, 3135 (2010).
\bibitem{Zheludev2012}J. Zhang, K. F MacDonald, and N. I. Zheludev, Light Sci. Appl. {\bf 1}, e18 (2012).
\bibitem{Huang2012}S. Huang, Z. Xie, W. Chen, J. Lei, F. Wang, K. Liu, and L. Li, Opt. Express {\bf 26}, 7066 (2018).
\bibitem{Salisbury1952}W. Salisbury, U.S. Patent No. 2,599,944, 1952.
\bibitem{Fante1988}R. L. Fante and M. T. McCormack, IEEE Trans. Antennas Propag. {\bf 36}, 1443 (1988).
\bibitem{Tischler2006}J. R. Tischler, M. S.  Bradley, and V. Bulovi\'{c}, Opt. Lett. {\bf 31}, 2045 (2006).
\bibitem{Piper2014}J. R. Piper and S. Fan, ACS Photonics, {\bf 1}, 347 (2014).
\bibitem{Fan2014}Y. Liu, A. Chadha, D. Zhao, J. R. Piper, Y. Jia, Y. Shuai, L. Menon, H. Yang, Z. Ma, S. Fan, F. Xia, and W. Zhou, Appl. Phys. Lett. {\bf 105}, 181105 (2014).
\bibitem{Nefedov2013}I. S. Nefedov, C. A. Valagiannopoulos, S. M. Hashemi, and E. I. Nefedov, Sci. Rep. 3, 2662 (2013).
\bibitem{Baranov2015}D. G. Baranov, J. H. Edgar, T. Hoffman, N. Bassim, and J. D. Caldwell, Phys. Rev. B {\bf 92}, 201405(R) (2015).
\bibitem{Inoue2002}H. Kato, T. Matsushita, A. Takayama, M. Egawa, K. Nishimura, and M. Inoue, IEEE Trans. Magn. {\bf 38}, 3246 (2002).
\bibitem{Liu2009}H. Liu, Z. Lou, H. Wang, and J. Miao, J. Infrared Millim. Terahertz Waves {\bf 30}, 401 (2009).
\bibitem{Shalaby2013}M. Shalaby, M. Peccianti, Y. Ozturk, and R. Morandotti, Nat. Commun. {\bf 4}, 1558 (2013).
\bibitem{Gurevich_book}A. G. Gurevich and G. A. Melkov, {\it Magnetization Oscillations and Waves} (CRC Press, Boca Raton, 1996).
\bibitem{Zvezdin_book}A. K. Zvezdin, and V. A. Kotov, {\it Modern Magnetooptics and Magnetooptical Materials} (Taylor \& Francis, New York, 1997).
\bibitem{Berning1957}P. H. Berning and A. F. Turner, J. Opt. Soc. Am. {\bf 47}, 230 (1957).
\bibitem{Makri2016}E. Makri, K. Smith, A. A. Chabanov, I. Vitebskiy, and T. Kottos, Sci. Rep. {\bf 6}, 22169 (2016).

\end{thebibliography}
\end{document}